\documentclass[twocolumn,showpacs,aps,prl,superscriptaddress]{revtex4}

\usepackage{graphicx}
\usepackage{amsmath}
\usepackage{latexsym}

\begin{document}

\title{Magnetic Field Induced Insulating Phases at Large $r_s$}

\author{G.A. Cs\'athy}
\affiliation{Department of
Electrical Engineering, Princeton University, Princeton, NJ 08544}

\author{Hwayong Noh}
\affiliation{Department of
Electrical Engineering, Princeton University, Princeton, NJ 08544}
\affiliation{Department of Physics, Sejong University, Seoul
143-747, Korea}

\author{D.C. Tsui}
\affiliation{Department of
Electrical Engineering, Princeton University, Princeton, NJ 08544}

\author{L.N. Pfeiffer}
\affiliation{Bell Labs, Lucent Technologies, Murray Hill, NJ 07974}

\author{K.W. West}
\affiliation{Bell Labs, Lucent Technologies, Murray Hill, NJ 07974}

\date{\today}

\begin{abstract}

Exploring a backgated low density two-dimensional hole sample in
the large $r_s$ regime we found a surprisingly rich phase diagram.
At the highest densities, beside the $\nu=1/3$, $2/3$, and $2/5$ 
fractional quantum Hall states, we observe both of the previously reported
high field insulating and reentrant insulating phases. 
As the density is lowered, the reentrant insulating phase
initially strengthens, then it unexpectedly starts weakening until it
completely dissapears.
At the lowest densities the terminal quantum Hall
state moves from $\nu=1/3$ to $\nu=1$.  
The intricate behavior of the insulating phases 
can be explained by a non-monotonic melting line
in the $\nu$-$r_s$ phase space.

\end{abstract}
\maketitle

A perpendicularly applied magnetic field ($B$) resolves the
density of states of a two-dimensionally confined charged system
into a set of discreet energies, called the Landau levels (LL).
When all carriers fall into the lowest LL
these systems have a variety of ground states.
The most prominent of these are the 
fractional quantum Hall liquids (FQHL) which develop at certain 
fractional values of the LL filling factor $\nu$
along with the Wigner solid (WS) \cite{fraq}. 
In the presence of disorder the condition for the 
development of the FQHL and the WS phases is thought to be
the dominance of the carrier-carrier interaction over
the carrier-disorder interaction.

The insulating phase at the highest $B$-fields found beyond 
$\nu=1/5$ in early measurements of
two-dimensional electron systems (2DES) has been
interpreted as a magnetically induced WS \cite{insul_e}. 
Numerous properties of this high field insulating phase (HFIP)
were found to be consistent with those of the WS \cite{review}. 
Of these, perhaps
the microwave resonances \cite{peide} 
constitute the most direct evidence that the HFIP is the long sought 
WS rather than a phase of singly localized particles.
The insulating behavior at low frequencies  
can be understood as the pinning of the WS by the impurities present 
in the host crystal and the microwave resonances are the
oscillation modes of the WS experiencing the pinning potential
of these impurities \cite{fukuyama}.
The HFIP has also been found in two-dimensional hole systems
(2DHS), but in the vicinity
of the terminal filling $\nu=1/3$ \cite{insul_h,rodgers,li}.
The difference between the 2DES and the 2DHS can be
qualitatively understood in terms of the evolution of the
boundary of the WS in the zero temperature limit, also referred to as the
melting line, in the $\nu$-$r_s$ phase space. 
The $r_s$ parameter has been introduced in metal physics as
a measure of the Coulomb interaction
in units of the Fermi energy and for a perfectly 2D system 
it is expressed as
$r_s = m^* e^2/(4 \pi \epsilon \hbar^2 \sqrt{\pi p})$, with $m^*$
being the effective mass and $p$ the areal density of the charges.
While to date the melting line lacks a full-fledged theory, several
of its properties are known. It is now well established that 
in the $r_s=0$ limit the melting line starts at $\nu \simeq 1/6.5$ 
\cite{eql,reentr}, it is believed that it monotonically moves
towards higher 
$\nu$ as $r_s$ increases \cite{reentr,platz,price,chui}, and it
extrapolates to  $r_s \simeq 37$ in the $B=0$ limit \cite{tanatar}.
Due to the larger effective mass of holes in GaAs-AlGaAs structures
at similar densities, $r_s$ is significantly larger than that of electrons
and, from the behavior of the melting line, a larger terminal filling results.

Simultaneously with the observation of the HFIP,
it was found that the insulating behavior persists even at $B$-fields
below that of the terminal FQHL, in particular in the range of 
$1/5<\nu<2/9$ in 2DES \cite{insul_e}
and $1/3<\nu<2/5$ in 2DHS \cite{insul_h}. This new insulating phase 
has been termed the reentrant insulating phase (RIP)
and it has also been interpreted as being the WS \cite{insul_e,insul_h}.
Thus the terminal FQHL is sandwiched in between
two WS phases, a property that is a result of the
lowering of the energy of the terminal FQHL below that of the WS
\cite{insul_e,insul_h}.
The RIP has been reported to strengthen
with decreasing density in 2DHS \cite{insul_h}. 
This property is consistent with 
that of the WS since at lower densities the WS is 
expected to be more stable \cite{reentr,platz,price,chui} resulting
therefore in a stronger insulating behavior.

Several features of the $\nu$-$r_s$ phase diagram remain unellucidated.
First, a recent study of a 2DHS having $r_s=30$ hinted that
there is no clear reentrant behavior at such large $r_s$ \cite{csa}.
This result indicates a reversal of the earlier
reported trend of strengthening of the WS with increasing
$r_s$ \cite{insul_h} and it implies an intricate evolution of the RIP.
Second, the shape of the melting line is unknown
at very large $r_s$ where the terminal quantum Hall state is
expected to move to a filling higher than $\nu=1/3$.
We note that these issues cannot be
tackled by current theories. One of the reasons is that
theories comparing the energies of the WS to that of
FQHL \cite{eql,reentr} make predictions at fractional
values of $\nu$ only, usually of the form of $1/m$ with $m$ 
an integer. In addition, the monotonic 
melting line obtained from a heuristic estimation
\cite{platz} and from a simple Lindemann-type criterion \cite{price,chui}, 
as well as a trivial density scaling of the energies of the WS
considered \cite{reentr} cannot account for the
collapse of the RIP hinted in Ref.\cite{csa}.

In this Letter we have systematically explored 
the evolution of the insulating phases, i.e. the
HFIP and the RIP, with density in the large $r_s$ limit.
To this end we performed standard low frequency magnetoresistance measurements
on a low density 2DHS that has been backgated. The holes are confined to
a 30~nm wide quantum well in a Si doped GaAs/AlGaAs heterostructure
grown on a (311){\it A} substrate. With the gate grounded the density 
is $p=1.63 \times 10^{10}$cm$^{-2}$ and the mobility
along the [$\bar{2}33$] axis at 33~mK
is $\mu=0.8 \times 10^{6}$cm$^2$/Vs.
The gate allowed a continuous tuning of the density from
$0.64$ to $2.85 \times 10^{10}$cm$^{-2}$. Using the recently reported
linealy increasing effective mass with density
\cite{keji}, the density range above corresponds to $r_s$ between 20 and 36.

\begin{figure}[!b]
\begin{center}
\includegraphics[width=3.3in,trim=0.0in 0.0in 0.0in -0.1in]{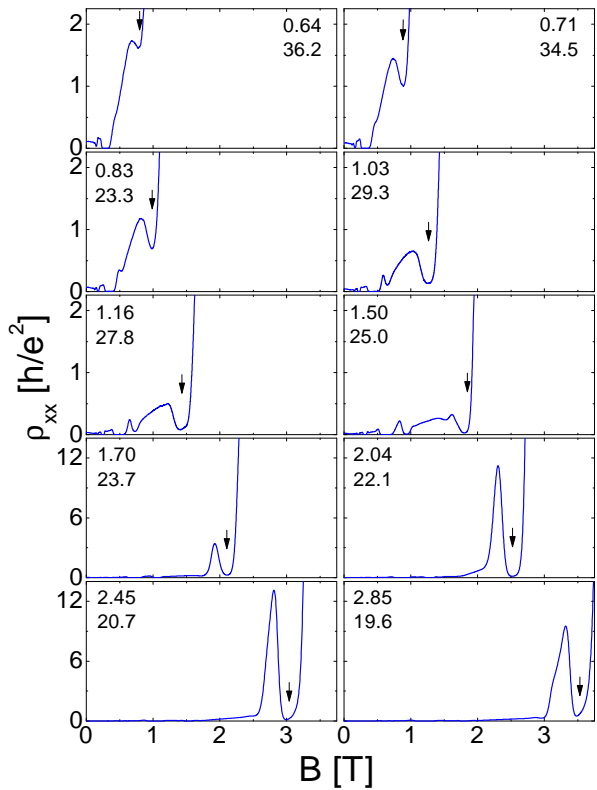}
\end{center}
\caption{\label{f1}
The longitudinal resistivity $\rho_{xx}$ 
in units of the quantum resistance
$h/e^2$ versus $B$ at 33~mK. Arrows mark $\nu=1/3$.
Each panel is labeled by the density in units of $10^{10}$cm$^{-2}$
(upper number) and the value of $r_s$ (lower number).
Note the change of the vertical scale for the 4 largest densities.
}
\end{figure}

In Fig.1 several traces of the diagonal resistivity $\rho_{xx}$
versus $B$ are shown for a set of representative
densities. The traces are taken at
the $T=33$~mK base temperature of our refrigerator. 
For all traces of Fig.1 we observe clearly developed $\nu=1$ and $2$
integer quantum Hall states and, with the exception of the lowest
densities, the $\nu=1/3$ and $2/3$ fractional quantum Hall states.
The steeply increasing $\rho_{xx}$ at the highest $B$ values signals the HFIP. 
At the highest density of $2.85 \times 10^{10}$cm$^{-2}$, similarly to the 
earlier studied 2DHS with larger densities 
\cite{insul_h}, our sample displays the reentrant insulating behavior. 
The peak in $\rho_{xx}$ at $B=3.3$~T
of peakheight $\rho_{xx}^{max} \simeq 9.5 h/e^2$, a value
that largely exceeds $h/e^2$, is due to this reentrant behavior.
As the density is decreased to
$2.45 \times 10^{10}$cm$^{-2}$, $\rho_{xx}^{max}$
increases. This property is similar to that of the
earlier measured 2DHS in the density range from 4 to 
$12 \times 10^{10}$cm$^{-2}$ \cite{insul_h} and it has
been interpreted as the strengthening of the WS phase
with decreasing density. 
With further lowering of the
density, however, the reentrant insulating peak
unexpectedly decreases and at $p=1.50 \times 10^{10}$cm$^{-2}$
is not present any more.
This non-monotonic dependence of $\rho_{xx}^{max}$
can better be seen in Fig.2a.
We find that $\rho_{xx}^{max}$ at 33~mK reaches its largest value
of $370$k$\Omega/\Box$ in the vicinity of 
$p=2.2 \times 10^{10}$cm$^{-2}$
and it vanishes at $p=1.54 \times 10^{10}$cm$^{-2}$.
At the intermediate densities of
1.50, 1.16, and $1.03 \times 10^{10}$cm$^{-2}$, while
the $\nu=1/3$ and $2/3$ FQHLs are well developed, the RIP
is clearly missing. These two FQHLs at the lowest densities 
of 0.83, 0.71, and $0.64 \times 10^{10}$cm$^{-2}$
show gradual weakening until they get buried in
a strong insulating background. 

\begin{figure}[!t]
\begin{center}
\includegraphics[width=2.6in,trim=0.0in 0.0in 0.0in 0.1in]{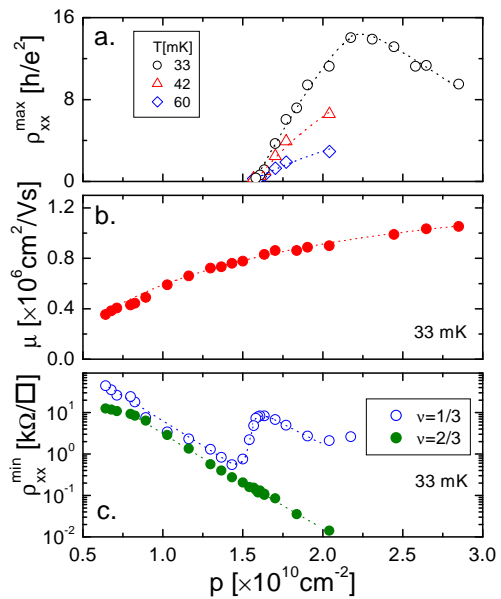}
\end{center}
\caption{\label{f2}
The dependence on density
of the peak resistivity $\rho_{xx}^{max}$ of the RIP 
at 33, 42, and 60~mK (panel a.),
the zero $B$-field mobility $\mu$ (panel b.), and the
resistivity minima $\rho_{xx}^{min}$ at $\nu=1/3$ and $2/3$ (panel c.)
at 33~mK. Lines are guides to the eye.
}
\end{figure}

We proceed now to the extraction of the
phase diagram in the $\nu$-$r_s$ phase space.
First we need to establish the values of the $B$-field 
at which the insulating phases lie.
One way to achieve this is to determine the crossing
of the $\rho_{xx}$ versus $B$ traces taken at different $T$ \cite{shahar}.
Our traces, however, do not have a clear crossing in
the $T$-range accessed (not shown).
In the absence of a critical $B$-field separating the
insulating and FQHL phases, we use the
$\rho_{xx} \geq h/e^2$ condition to delimit the insulating phases.
Such a definition is consistent with the
$T$-dependence of the traces we measure.
Furthermore we note that a small error in determining
the exact location of the onset of the
insulating behavior has only a minor effect;
we expect that the qualitative features of the phase diagram 
to be insensitive to the relaxation of the condition above.
By performing the described procedure, we obtain the
phase diagram shown in Fig.3. In addition to our data (circles), 
Fig.3 is completed at $r_s<20$ with data
from an earlier measurement of ours of a 2DHS with $r_s=18.5$,
from Ref.\cite{insul_h} (squares), and from Ref.\cite{rodgers} (triangles).

\begin{figure}[b]
\begin{center}
\includegraphics[width=3.55in]{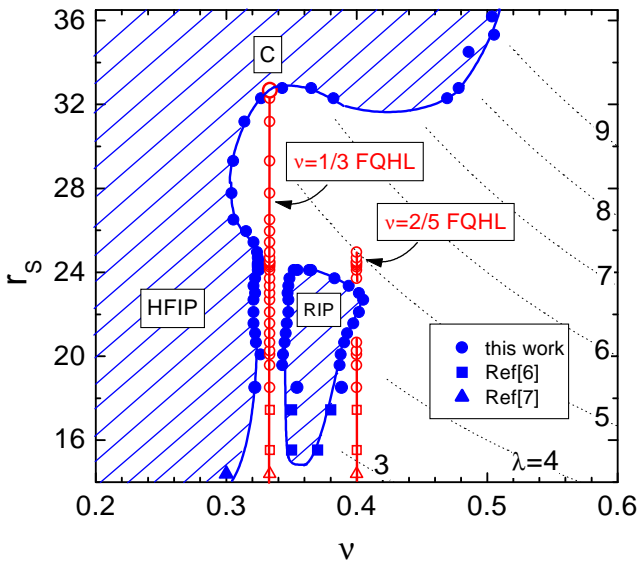}
\end{center}
\caption{\label{f4}
The boundary of the insulating phases (full symbols) and
the $\nu=1/3$ and $2/5$ FQHL (open symbols) 
in the $\nu$-$r_s$ phase space. Continuous lines are guides
to the eye. Dotted lines are loci of points obeying
$E_c = \lambda \cdot \hbar \omega_c$ (see text).
}
\end{figure}

The obtained phase diagram is unexpectedly rich.
Besides the FQHL at $\nu=1/3$ and $2/5$, the phase diagram
is dominated by a tortuous boundary of the HFIP and RIP.
For $r_s<24$ both the RIP and the HFIP are present and
in between them lies the $\nu=1/3$ FQHL.
The $\nu=2/5$ FQHL extending up to $r_s \simeq 25$
appears to be interrupted by the RIP.
In the $20<r_s<22$ range the boundary of the RIP on the higher $\nu$ side
is described by an increasing $r_s$ with increasing $\nu$. 
This trend is consistent with the increase of the
terminal filling with $r_s$
from 1/5 in 2DES to 1/3 in 2DHS \cite{insul_e,insul_h},
with the earlier reported density
dependence of the RIP \cite{insul_h}, and with theoretical
expectations. Indeed, predictions of the melting line 
using heuristic arguments \cite{platz}, from a simple
Lindemann-type of melting criterion \cite{price,chui}, as well as from a more
sophisticated incorporation of the mixing of the LL \cite{reentr,price} yield
an increasing $r_s$ with $\nu$. 
We note that at the lowest $r_s$ values of Fig.3
the RIP is not present \cite{insul_h,rodgers}.
For $r_s$ between 22 and 24 
we find a decreasing $r_s$ as a function of increasing
$\nu$ and the RIP surprisingly dissapears for $r_s>24$.
The collapse of the RIP correlates with the vanishing $\rho_{xx}^{max}$
of Fig.2a. This collapse is anomalous as it cannot be accounted
for by existing theories. We note that the dissappearance of the RIP
close to $r_s=24$ is not a finite $T$ effect. Indeed, Fig.2a shows
that the density or the $r_s$ of this collapse is independent of $T$ 
and, furthermore, for $r_s>24$ we observe 
neither the RIP nor any precursor of it.

In the $24<r_s \lesssim 32$ region of the phase diagram, while
the RIP is absent, the HFIP and the $\nu=1/3$
FQHL are well developed. At even larger $r_s$, there is a sliver
of the phase diagram ranging from $r_s \simeq 32$ to $33$ with quite
interesting properties. Here the $\nu=1/3$ FQHL, which is expected
to fully develop in the $T=0$ limit, is intercalated between
two insulating phases. The insulating phase at the higher
filling factor, however, does not resemble the RIP present at
$r_s<24$. The difference is that this new insulating phase
extends to much larger $\nu$, closer to $\nu=1$, 
and the height and the $T$-dependence of the resistivity peak are both
smaller than that of the RIP. At 
$r_s>33$, the $\nu=1/3$ FQHL cannot be supported any more
and the terminal quantum Hall state moves to the integer value of $\nu=1$.
Such a $\rho_{xx}$ versus $B$ trace is shown in the first panel of Fig.1
where $p=0.64 \times 10^{10}$cm$^{-2}$.

From the phase diagram in the $\nu$-$r_s$ plane it 
seems that the RIP forms an island that is
topologically disconnected from the HFIP. However,
at a more careful inspection it becomes apparent that
the phase boundary of the RIP that is not parallel
to the $\nu=1/3$ vertical line is a prolongment
of the boundary of the HFIP. In other words, one can
understand the phase diagram as originating from a single
insulating phase that is bound by a continuous and non-monotonic
S-shaped melting line spanning the entire $r_s$ range. 
This insulating phase is interrupted
at $\nu=1/3$ by the FQHL. In this interpretation, the RIP and the
HFIP have the same origin. The melting of the WS at the
S-shaped boundary is attributed to
quantum fluctuations caused by the zero point motion \cite{review,platz,chui}
while melting at the two boundaries parallel to the $\nu=1/3$ vertical line, 
i.e. the phase boundary of the HFIP and that of the RIP
on the lower $\nu$ side, is due to Laughlin correlations \cite{melt2}.
We note that small variations of the phase digram are expected
to occur in samples of various disorder and finite layer 
thickness \cite{insul_h} without a change in the topology of the phase diagram.

In the following we discuss four 
possibilities for the collapse of the RIP with decreasing
density. The first such possibility is related to the
degradation of the sample quality as the density is lowered.
Indeed, as shown in Fig.2b, 
the mobility $\mu$ of the 2DHS decreases with decreasing density.
In an extreme case it is possible that the WS
cannot be supported any more and a new
insulating phase sets in. This new insulating phase could be 
another collective insulator or, more likely, a phase of
singly localized particles. At the collapse of the RIP
in the vicinity of $p=1.54 \times 10^{10}$cm$^{-2}$, however,
the mobility has a very gentle variation and $\rho_{xx}^{max} \ll h/e^2$. 
We therefore think that such a scenario is unlikely. A second possibility is 
a periodic modulation of the melting line imposed by LL mixing.
Such a modulation could arise every time the
Coulomb energy $E_c$ encompasses an integer number of Landau levels
$\hbar \omega_c$, i.e. $E_c=\lambda \cdot \hbar \omega_c$
\cite{yoshioka}. The loci of points obeying the condition above are
plotted on the phase diagram for several values of $\lambda$. Since
the anomalous behavior of the melting line does not possess any
periodicity as $\lambda$ is incremented, we conclude that the effect
described above is not a good explanation either.
The third scenario involves a commensuration effect that is
similar to the registry of Helium atoms with a 
graphite substrate \cite{graphite} . If 
the potential the holes experience has a periodic spatial modulation
of areal density of the minima of $2.2 \times 10^{10}$cm$^{-2}$, 
this potential will favor
the WS of the same density and the melting line will protrude to
higher $\nu$. Since we cannot identify any periodicity of such a lengthscale
in the host GaAs crystal, we discard this possibility as well.
Finally, the fourth scenario relates to the composite-fermion WS interpretation
of the insulating phases \cite{fertig}. The Laughlin-Jastrow correlated
trial wavefunction used is built on the single particle wavefunctions
of the lowest LL. We note, however, that the RIP is beyond the
lowest LL of the composite-fermions. If the wavefunction of the
second LL is included with a weight, it could become energetically
more favorable for the width of the wavefunction
to spread out with decreasing density. We propose that the collapse
of the RIP could be due to such a relaxation of the single particle
wavefunction. Due to the spreading of the wavefunction there is less
distance between the lattice sites of WS. This can lead to
an increased zero point motion that melts the WS. 

It is noteworthy to point out the behavior
of the $\nu=1/3$ FQHL near the collapse of the RIP at
$p=1.54 \times 10^{10}$cm$^{-2}$.
As the the density is increased, at constant $T$ it is expected that FQHLs 
develop deeper resistivity minima $\rho_{xx}^{min}$.  
While $\rho_{xx}^{min}$ at $\nu=2/3$ does have such a behavior, 
at $\nu=1/3$ $\rho_{xx}^{min}$ has an unexpected
increase between 1.47 and $1.62 \times 10^{10}$cm$^{-2}$
(shown in Fig.2c). Since this anomalous behavior develops 
at the crossing of the
S-shaped melting line described earlier and the $\nu=1/3$ line, 
we think that it is a signature of the quantum fluctuations
of the zero-point motion affecting the FQHL.

Our data implies, that the $r_s \simeq 33$ and $\nu=1/3$ point is special.
This point, denoted by $C$,
is the endpoint of the $\nu=1/3$ FQHL line and it is on the melting
line of the solid as well. 
Thus, in this point there is an intriguing possibility 
of coexistence of the FQHL and the WS phases.
This region of the phase diagram has recently been accessed
in a slighly lower quality 2DHS \cite{csa} and the observed
periodic modulation of $\rho_{xx}$ has been attributed to such
a two-phase coexistence.

In conclusion, we have presented a thorough exploration of the
phase diagram in the $\nu$-$r_s$ plane of two-dimensional
holes in the lowest LL. 
We think that this diagram
can be understood as originating from a monolithic insulating phase 
that is cut by the $\nu=1/3$ FQHL into two disjoint phases: the
HFIF and the RIP. The RIP appears to form an isolated island
because of the non-monotonic shape of the melting line. 

We thank D. Huse, M. Manfra, and C. Zhou for their comments.
This research was funded by the DOE and the NSF.

\end{document}